\documentclass[12pt,a4paper]{article}
\usepackage[T1]{fontenc}
\usepackage{dfttrob}
\usepackage[numbers,sort&compress]{natbib}
\usepackage{latexuseful2e}
\usepackage{amsmath}
\usepackage{graphicx}

\dfttnum{FISIST/14-2000/CENTRA\\September 2000}

\newcommand{\nuNA}{\nu_{{}_{N_A}}}
\newcommand{\SNA}{S_{{}_{N_A}}}
\newcommand{\Jpsi}{J/\psi}                
\newcommand{\Npart}{N_{\text{part}}}    

\begin{document}

\title{Particle Rapidity Density Saturation in Heavy Ion Collisions and 
the Dual String Model}
\author{J. Dias de Deus and R. Ugoccioni\\
 \it CENTRA and Departamento de F{\'\i}sica (I.S.T.),\\ 
 \it Av. Rovisco Pais, 1049-001 Lisboa, Portugal}
\maketitle

\begin{abstract}
We show that the Dual String Model with fusion leads, in
heavy ion collisions, to strict saturation of the particle
(pseudo-)rapidity density, normalised to the number of participant
nucleons, as that number increases. Asymptotically, as $\sqrt{s} \to
\infty$, with the number of participants fixed, this density
approaches the nucleon-nucleon density.

A comparison with recent WA98 data is presented.
\end{abstract}

\section{Introduction}

Recently, experimental \cite{phobos:1,WA98:1,Lourenco} and 
theoretical \cite{Wang,RU:dNdeta} papers have
addressed the question of the dependence of measurable quantities
(charged particle density, transverse energy, $\Jpsi$ production rate)
on the number $\Npart$ of participant nucleons in high energy heavy
ion collisions.
That information, as stressed in \cite{Wang}, 
as well as in \cite{RU:dNdeta}, is
extremely important as it allows for a better understanding of the
initial conditions in the evolution of newly created dense matter and
provides the information for discriminating among different models.

In the analysis of the data of \cite{phobos:1} made in \cite{Wang} it was
emphasised the role of the dependence of the particle rapidity
density, $dN/dy$, normalised to the number of pairs of participants,
and two models were considered.

The HIJING Monte Carlo model \cite{hijing} contains soft and semi-hard
interactions in an unitarised form, the $p_T$ threshold for jet
production, $p_0$, being independent of energy and atomic mass number
$A$.
As in the model there is a growing number of hard collisions, growing
faster than $\Npart$, the quantity
\begin{equation}
	\Phi(\Npart,y,\sqrt{s}) \equiv \frac{1}{\frac{1}{2}\Npart} \frac{dN}{dy}
																										\label{eq:1}
\end{equation}
is an increasing function of $\Npart$, or
\begin{equation}
	\frac{\partial \Phi}{\partial \Npart} > 0  .			\label{eq:2}
\end{equation}

The EKRT model \cite{EKRT}, on the other hand, relates the hard physics
threshold $p_s$ to a saturation criterion \cite{saturation}: the number of
partons multiplied by their effective area ($\approx 1/p_T^2$) must be
less than the transverse area of interaction.
The saturation transverse momentum $p_s$, in contrast to the parameter
$p_0$ of the previous model, is a growing function of $\sqrt{s}$ and
$A$.
The assumption that physics is controlled by $p_T = p_s$ leads to the
result that $\Phi$ is a {\em decreasing} function of $\Npart$,
\begin{equation}
	\frac{\partial \Phi}{\partial \Npart} < 0   .			\label{eq:3}
\end{equation}

\begin{figure}
  \begin{center}
  \mbox{\includegraphics[width=0.62\textwidth]{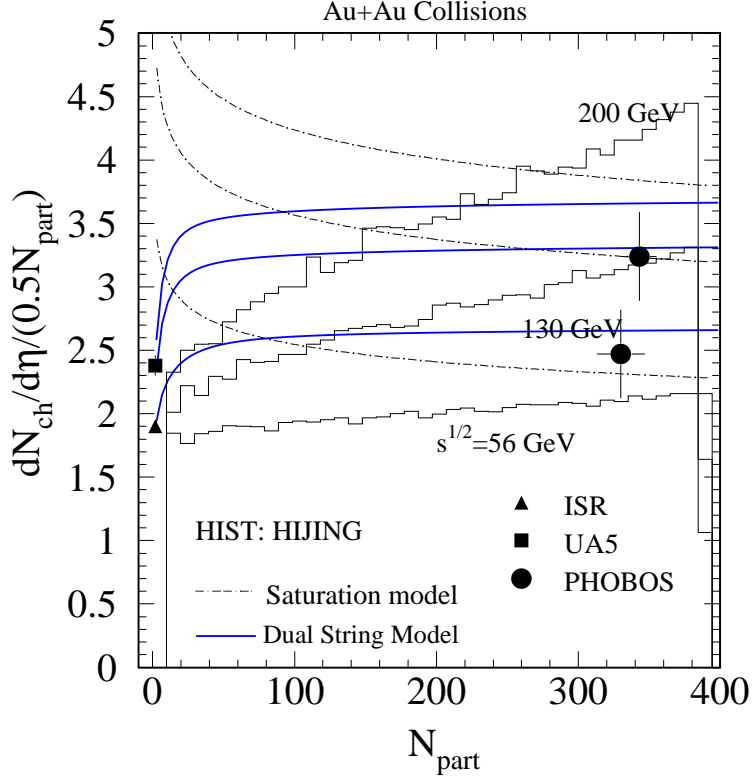}}
  \end{center}
  \caption[Central charged particle rapidity density per 
	participating pair]{Central charged particle rapidity density 
	per participating pair as
  a function of the number of participants. Results of HIJING
  (histograms), EKRT predictions (dot-dashed lines) and DSM predictions
  (solid lines) for central Au+Au collisions at $\sqrt{s}=56,130,200$
  $A$GeV. Also shown are results from $pp$ and $p\bar p$ collisions
  and PHOBOS data (Everything in the figure except the DSM curves is 
  taken from \cite{Wang}.)}\label{fig:wang}
  \end{figure}

In Fig.~\ref{fig:wang}, taken from \cite{Wang}, 
we show the predictions for the two
models, histogram for HIJING and dot-dashed line for EKRT.

The Dual String Model (DSM) with string fusion was developed in
\cite{RU:dNdeta}.
Concerning this model two remarks can be made:

\textit{(i)} There exist two components, a valence-valence component,
proportional to $\Npart$, and a sea-sea (including gluons) component
which grows faster than $\Npart$ and corresponds to multiple scattering.
We have a structure somewhat similar to the HIJING model.
Without fusion DSM gives an increase of $\Phi$ as, roughly, $\Npart^{1/3}$.

\textit{(ii)} With fusion, DSM leads to saturation of $\Phi$ as
$\Npart$ increases, not very different from what is obtained in the
EKRT model (see, in particular, \cite{EKRT:2}).

In conclusion, in DSM with fusion,
\begin{equation}
	\frac{\partial \Phi}{\partial \Npart} > 0  ,			\label{eq:4}
\end{equation}
as in HIJING, and, as $\Npart$ becomes larger and larger,
\begin{equation}
	\frac{\partial \Phi}{\partial \Npart}
	\xrightarrow[\Npart\to\infty]{} 0  ,							\label{eq:5}
\end{equation}

The $\Npart$ dependence of $\Phi$ in DSM with fusion is also shown in
Fig.~\ref{fig:wang} (solid lines). The parameters used are precisely
those used in \cite{RU:dNdeta} 
to describe NA49 and PHOBOS data \cite{phobos:1}.

\section{DSM and particle densities}

DSM is essentially the Dual Parton Model \cite{DPM:1} 
with the inclusion of strings \cite{StringFusionModel}.
The strings may interact by fusing 
\cite{Pajares:perc+NardiSatz+jotapsi} in the
transverse plane of interaction and eventually one may reach a
situation of percolation with the formation of extended regions of
colour freedom, with the features of the expected Quark-Gluon Plasma.

In the Dual Parton Model hadrons are considered as made up of
constituents quarks (valence and sea quarks) and gluons.
Two basic diagrams contribute to particle production (see \cite{RU:dNdeta}).
The valence-valence diagram corresponds to
single inelastic scattering and the wounded nucleon model \cite{WNM}.
The sea-sea diagram (including gluons)
corresponds to the additional inelastic multiple scattering
contributions. 
These contributions may be internal, parton multiple scattering within
the original valence-valence contribution, or external, involving other
nucleons. 

If $h$ is the height of the valence-valence plateau, $\alpha h$ the
height of the sea-sea plateau, $2k$ the average number of strings
produced in a nucleon-nucleon collision we can write
\cite{Armesto:SFM,RU:dNdeta} for the (pseudo-)rapidity particle
density
\begin{equation}
	\left.\frac{dN}{dy}\right|_{N_AN_A} = 
		N_A \left[ 2 + (2k-1)\alpha\right] h
				+ (\nuNA - N_A)  2k\alpha h ,	
																			\label{eq:Armesto}
\end{equation}
where $N_A$ is the number of pairs of participant nucleons
\begin{equation}
	N_A = \frac{1}{2} \Npart    , 				\label{eq:7}
\end{equation}
and $\nuNA$ is the average number of nucleon-nucleon collisions. From
elementary multiple scattering arguments \cite{Armesto:SFM} we have
\begin{equation}
	\nuNA = N_A^{4/3}							\label{eq:nu_NA}
\end{equation}

One should notice that the number of nucleon-nucleon collisions is
$N_A+(\nuNA-N_A) = \nuNA$ and the number of strings is $N_A\left[ 2
+2(k-1)\right] + (\nuNA - N_A) 2k = 2k\nuNA $.

For $N_A=1$, nucleon-nucleon collision, Eq.~(\ref{eq:Armesto}) gives
\begin{equation}
		\left.\frac{dN}{dy}\right|_{pp} = \left[ 2 +2(k-1)\alpha \right] h   .
																\label{eq:pp}
\end{equation}

Assuming that $h$ and $\alpha$ are energy independent (constant
plateaus) the energy dependence of ${dN}/{dy}|_{pp}$ fixes the energy
dependence of $k$.

From Eq.s~(\ref{eq:1}), (\ref{eq:Armesto}), (\ref{eq:7}),
(\ref{eq:nu_NA}) and (\ref{eq:pp}) we can write
\begin{equation}
	\Phi(N_A, \sqrt{s}, y) = \phi(\sqrt{s}, y) + (N_A^{1/3} - 1) 2 k
	\alpha h																				\label{eq:10}
\end{equation}
with
\begin{equation}
	\phi(\sqrt{s}, y) \equiv \left.\frac{dN}{dy}\right|_{pp}    .
\end{equation}

The number of strings coming from nucleon multiple scattering ---the
second term in Eq.\ (\ref{eq:Armesto})---is $N_A (N_A^{1/3}-1) 2 k$ and
they occupy the transverse interaction area $\SNA$, which, for central
collisions, is approximately given by
\begin{equation}
		\SNA \simeq \pi \left(1.14 N_A^{1/3}\right)^2   ,		\label{eq:12}
\end{equation}
such that the dimensionless transverse density parameter $\eta$ is
\begin{equation}
	\eta = \left( \frac{r_s}{1.14} \right)^2 2 k N_A^{1/3} 
					(N_A^{1/3} - 1) ,														\label{eq:13}
\end{equation}
where $r_s \simeq 0.2$ fm is the string transverse section
radius. Note that $\eta$ increases with $N_A$ and $\sqrt{s}$.

When fusion occurs, i.e., when strings cluster in the transverse
plane, Eq.\ (\ref{eq:10}) becomes \cite{RU:dNdeta}
\begin{equation}
	\Phi(N_A, \sqrt{s}, y) = \phi(\sqrt{s}, y) + F(\eta) (N_A^{1/3} - 1) 2 k
	\alpha h	,																			\label{eq:14}
\end{equation}
where $F(\eta)$ is the particle production reduction factor
\cite{Braun:Feta},
\begin{equation}
	F(\eta) \simeq \sqrt{ \frac{1-e^{-\eta}}{\eta} }	.		\label{eq:15}
\end{equation}

We shall next discuss the $N_A$, $\sqrt{s}$ and $y$ dependence of
$\Phi$, Eq.\ (\ref{eq:14}).

For the $N_A$ dependence we have
\begin{equation}
	\frac{\partial \Phi}{\partial N_A^{1/3}} = 2 k \alpha h 
			\frac{\partial}{\partial N_A^{1/3}} \left[ F(\eta) (N_A^{1/3} -
			1) \right] ,
\end{equation}
and, making use of the definition of $\eta$, Eq.\ (\ref{eq:13}),
\begin{equation}
	\frac{\partial \Phi}{\partial N_A^{1/3}} = 
		\frac{1.14}{r_s} \sqrt{ 2 k } \alpha h 
		\frac{\partial}{\partial N_A^{1/3}} \left[ (1-e^{-\eta})
			\frac{N_A^{1/3} - 1}{N_A^{1/3}} \right]^{1/2} ,   \label{eq:16}
\end{equation}
As both $(1-e^{-\eta})$ and $(N_A^{1/3} - 1)/N_A^{1/3}$ are increasing
functions of $N_A$, becoming constant as $N_A\to\infty$,
\begin{equation}
	\frac{\partial \Phi}{\partial N_A} > 0 \qquad\text{with}\qquad
	\frac{\partial \Phi}{\partial N_A} \to 0 \quad\text{as}\quad N_A \to\infty .
																						\label{eq:18}
\end{equation}
These are the results announced in the Introduction, Eq.s (\ref{eq:4})
and (\ref{eq:5}).

The DSM with fusion thus predicts {\em saturation of the particle rapidity
densities per participant pair of nucleons, $\Phi$, as $N_A$ increases}. 

\begin{figure}
  \begin{center}
  \mbox{\includegraphics[width=\textwidth]{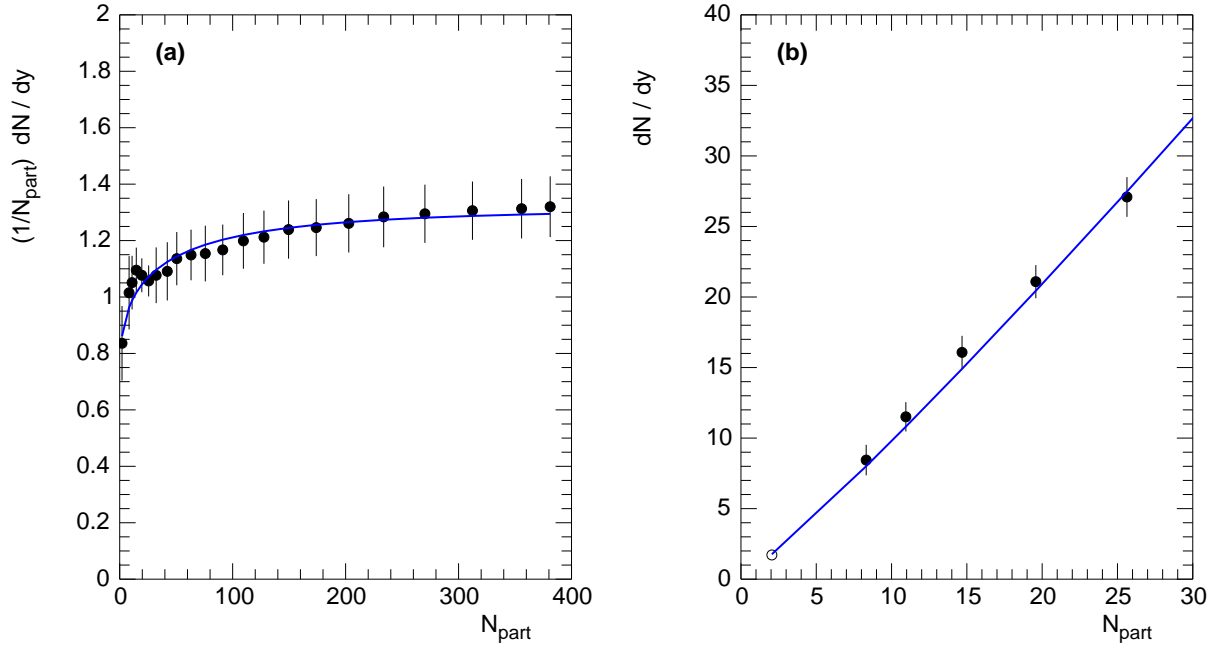}}
  \end{center}
  \caption[Charged particle density]{\textbf{(a)} Charged particle density per 
	participant nucleon versus the number of participants;
	\textbf{(b)} absolute charged particle density
	versus the number of participants. The data from WA98 \cite{WA98:1}
	refer to $158 A$ GeV Pb+Pb collisions (filled circles), the open
	circle refers to $pp$ collisions \cite{DeMarzo}; the solid line
	results from Eq.\ (\ref{eq:14}). }\label{fig:wa98}
  \end{figure}

In Fig \ref{fig:wa98} we compare our model Eq.\ (\ref{eq:14}) with
the very recent results \cite{WA98:1} of WA98 Collaboration on central Pb+Pb
collisions at the CERN SPS. For large values of $\Npart$ the
saturation is very clearly seen, Fig.\ \ref{fig:wa98}a, while for small
values of $\Npart$ the rise of particle density with $\Npart$, 
Fig.\ \ref{fig:wa98}b, is well described.
The values used for the parameters were $h = 0.77$ and $\alpha = 0.11$,
and for the $pp$ density we used the same value as WA98, from
\cite{DeMarzo}.
Note that the parameters here have slightly different values in
comparison to the ones used in \cite{RU:dNdeta}.
This is not surprising as the WA98 data are
not compatible with NA49 data as given by PHOBOS \cite{phobos:1}.

Concerning the energy dependence of Eq.\ (\ref{eq:14}), if the energy
is high enough
\begin{equation}
	\phi(\sqrt{s}, y) \equiv  \left.\frac{dN}{dy}\right|_{pp} \sim 2 k
	\alpha h  ,
\end{equation}
and
\begin{equation}
	\Phi(N_A, \sqrt{s}, y) / \phi(\sqrt{s}, y) = 
			1 + F(\eta) (N_A^{1/3} - 1)   .			\label{eq:21}
\end{equation}
As $F(\eta)$ goes to zero as $\sqrt{s}$ increases we obtain the
prediction that {\em asymptotically the particle density per
participant, for a fixed number of participants, approaches the
nucleon-nucleon density and this approach is controlled by the
function $F(\eta)$}, Eq.\ (\ref{eq:15}).
This means that asymptotically heavy ion collisions, with respect to
particle densities, become similar to nucleon-nucleon collisions. 
This is a consequence of the increasing role played by string fusion.

\begin{figure}
  \begin{center}
  \mbox{\includegraphics[width=0.5\textwidth]{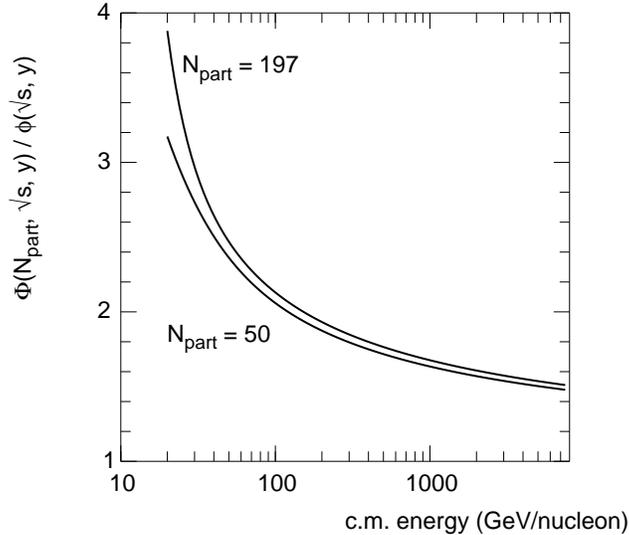}}
  \end{center}
  \caption{Ratio of the particle rapidity density per participating
  pair to the $pp$ particle density, Eq.~(\ref{eq:21}), as function
	of c.m.\ energy for fixed values of the number of participants,
	$\Npart$.}\label{fig:3}
  \end{figure}

In Fig.\ \ref{fig:3} we present our prediction for the dependence of
the ratio (\ref{eq:21}) on the energy for fixed values of $\Npart$.

Regarding the (pseudo-)rapidity dependence of $\Phi$,
Eq.\ (\ref{eq:14}), even without a more specific model, we can say that
the valence-valence contribution dominates the large centre-of-mass
$|y|$ region, as most of rapidity is taken by the valence diquarks.
{\em In the fragmentation region we expect the particle density per
participant nucleon to be equal to the nucleon-nucleon density,}
\begin{equation}
	\Phi(N_A, \sqrt{s}, y_{\text{fragm}}) \simeq 
		\phi(\sqrt{s}, y_{\text{fragm}})  .
\end{equation}
The sea-sea contributions are naturally shorter and more central in
rapidity.

\section{Conclusions}

The DSM is a model with two components, the valence-valence component
and the sea-sea component, the sea-sea component increasing its
importance with energy and number of participants.
This is somewhat similar to the HIJING Monte Carlo model, with soft
and hard components.

On the other hand, with fusion the DSM behaves, for large $N_A$,
similarly to the EKRT model, but with strict saturation of the
particle density per participant nucleon.
However, in the original EKRT model the saturation criterion in the
transverse plane is stronger than in case of fusion of strings.
Here, saturation in the interaction area is asymptotic (when
$\eta\to\infty$) while in the EKRT model it occurs, using the string
language, when $\eta \equiv (\pi r_s^2) / (\pi 1.14^2 N_A^{2/3}) (2 k
N_A (N_A^{1/3} - 1)) = 1$.
This causes the decrease of $\Phi$ with $N_A$ in the EKRT original
model.

Probably different explanations, such as the ones based on string
fusion, parton saturation, parton shadowing, are in some sense dual
and refer to the same underlying physics \cite{Armesto:SFM}.
What is becoming clear is that saturation of particle density puts
strong constraints in models, and limits the rise of the
(pseudo-)rapidity plateau at RHIC and LHC.

\section*{Acknowledgements}
R.U. gratefully acknowledges the financial support of the
Fundação Ciência e Tecnologia via the ``Sub-Programa Ci\^encia 
e Tecnologia do $2^o$ Quadro Comunit\'ario de Apoio.''

\bibliographystyle{prstyR}  
\bibliography{abbrevs,bibliography}

\end{document}